# The Silicon photomultiplier as a metasystem with designed electronics as metadevice for a new receiver-emitter in visible light communications

R.M. Gutierrez*[a], A.I. Hernandez[a], L.F. Castañeda[a], J.F. Castaño[a]
[a]Centro de Investigación en Ciencias Básicas y Aplicadas (CICBA), Universidad Antonio Nariño
Bogotá, Colombia

## ABSTRACT

A Silicon Photomultiplier, SiPM, is a metasystem of Avalanche Photodiodes, APDs, which embedded in a specific purpose electronic, becomes a metadevice with unique and useful advanced functionalities to capture, transmit and analyze information with increased efficiency and security. The SiPM is a very small state of the art photo-detector with very high efficiency and sensitivity, with good response to controlled light pulses in the presence of background light without saturation. New results profit of such metadevice to propose a new receiver-emitter system useful for Visible Light Communication, VLC.

## 1. INTRODUCTION

Visible Light Communication (VLC), is a recent field of application in the area of optical communications with intense research activity [1, 2, 3]. An important research topic in this area, is the development of an adequate system of reception and discrimination of the light signals, which must have a high efficiency, low cost and broad applicability. VLC communication systems are standardized by IEEE 802.15.7 standard of 2011[4]. This standard defines aspects such as basic type modulation as flickering **, transfer rates, techniques of error detection and correction, intensity management of the transmitters by dimmering ***, it proposes schemes of access control, synchronization, authentication of users and some other technical aspects, all integrated within the overall context of the IEEE 802.15 standards specifying the conditions of operation of wireless communication systems in General[5]. The evolution of the communication technology has kept under continuous review, especially regarding transfer rates, modulation techniques and the inclusion of other advanced technical aspects of VLC networks such as the spatial distribution of the network elements, convergence with other networks, the use of infrared and near ultraviolet light, and specific applications among others. These activities are coordinated by the IEEE Working Group for Wireless Personal Area Network for all the set of standards 802.15, planning a draft version of the standard 802.15.7 by the end of 2016 and the final version for the year 2017[6].

In VLC visible light transports the information and therefore it has to be emitted and received by modulation in digital or analog form, similarly to conventional systems of radio frequency (RF), replacing traditional emitter-receivers by optical emitter-receivers. Therefore, VLC needs an electronic system including photodetectors and signal conditioning circuits with the ability to discriminate and reconstruct the signals of interest. One of the major challenges of VLC is to be able to operate in background light conditions (sunlight and other lighting systems). These challenges means the implementation of a systems of high-speed and precision overcoming the natural limitations that light may have for communications such as dispersion and the difficulty to overcome physical obstacles. However, as we will show, these limitations can be transformed into advantages in certain applications offering particular security for the information and lower interference with surrounding electronics. In other applications, these features of VLC can also reduce health risks for users given that the low intensities of visible light characteristic of VLC, are harmless for humans and most forms of live and other sensitive nano and micro structures such as electronic equipment in hospitals, airplanes, industrial plants, etc. Thus VLC can be an excellent communication system in specific environments coexisting and using the infrastructure of traditional and emergent artificial lighting and technology using for example LED (Light Emitter Diode) emerging[7].

---

[*]director.sistemas.complejos@uan.edu.co
**flickering: rapid variation on intensity, normally on/off conmutation
***dimmering: control of light intensity

Currently, there are four possible implementations of VLC systems, similar to those based on radiofrequency WiFi, considering transmission capacity and network VLC[8] target systems. In the implementation Docking, the devices are close and controlled by a fixed structure; for the system Beaming, one of the devices is mobile and directionally used on another device; for the implementation Hotspot, there is a central element of transmission-reception for emission to directionally connected devices; for the Broadcasting system, the transmission and reception is omnidirectional and the devices are in transmission range with background radiation. The last system may be implemented as a full-duplex communication multi-point system consisting of simultaneous communication between multiple users through a hub. This last type of VLC communication system is the most demanding in all aspects, in both the computational requirements and the characteristics of the optical receivers-transmitters, requiring the implementation of multiple access techniques such as Orthogonal Frequency Division Multiplexing (OFDM), or the like, which is an intensive subject of research. [9, 10, 11]

At the Association for Computing Machinery (ACM), Workshop 2014, it was presented a comprehensive review of the state of the art in communications with visible light, the current challenges and research trends that have appeared worldwide, showing that there is huge potential in the development of VLC in different scenarios, giving a framework to address inherent problems and find solutions to the advancement of such tecnologías[12]. One of the main challenges that raised in this event, proposed as main subject for current research, is the improvement of the optical receivers so they can operate at lower light intensities, with higher transfer rates and capable to operate at high backlight intensity without the use of optical filters, with the versatility for applications in places like streets, open spaces, parks and also special closed ambiences as those mentioned above.

This article describes a technological solution of a VLC transmitter-receiver based on silicon photo multipliers (SiPM), and an appropriate electronic system that can work very well with high background radiation. Specific and very special features of this modern photo detector, the SiPM, is used to implement a VLC receiver operating in backlight conditions, with higher transfer rates and requiring less light intensity. Implementing the SiPM and the electronic associated into the area of VLC, is a new proposal that may help to solve the most important problems of VLC to become a real and efficient system of communications with new advantages.

This article is organized as follows: in section 2 it is presented a description of VLC, its problems and limitations; in Section 3 we present a detailed description of the installation and experimental process that was implemented to characterize the VLC transmitter-receiver metadevice with the SiPM as Metasystem that, to our knowledge, have not been proposed so far, presenting also how this new approach contributes to the solution of the issues raised in VLC; in Section 4 we present and analyze the results of the experimental process, the VLC SiPM based transmitter-receiver metadevice developed and its conditions of operation in the context of VLC .

## 2. PROBLEM DESCRIPTION

So far, VLC has limitations related to the transfer rate, transmission mode, ambient operating conditions, transmission conditions, operating distance and modulation technique, as well regarding the dimmering of the transmission signal (light).[13] The Institute Fraunhofer has a commercial implementation of VLC that reaches a transfer rate of 3Gbps and a reduced version that reaches 500 Mbps[14] with a maximum operating distance of 4 m. This is a unidirectional system, which uses RGB LED white light, not a conventional lighting system; additionally it requires the use of optical filters to avoid the influence of sunlight or conventional lighting. The University of Edinburgh and the company pureLiFi[TM] are developing the system Li-Flame[TM]; a first prototype was presented in March 2015[15] and it reaches a transfer rate of 10 Mbps, allowing two-way communication (full-duplex) between different users by using an upstream channel with a channel for infrared light and a down channel for white light from conventional LEDs, the operating distance is 3 m; this system requires the use of optical filters to discriminate the communication signal from background. The background is a natural condition that has to be considered for VLC, it comes from different light sources that not always can be separated from the light source of information transmission, like lighting lamps, sunlight and any other source of visible light within the spectral range of the receiver[16].

Therefore, VLC focuses on 6 main challenges: 1. achieve transfer rates of hundreds of Mbps, 2. use a light source coexisting with the lighting source without introducing disturbances to light perception of users, 3. operate in spatial ranges typical of housing and labor premises (houses and buildings, offices, meeting rooms, etc.), 4. Employ efficient control of lighting or dimmering without requiring the use of optical filters as discriminators of background and signal 5. Enable bidirectional connection of multiple users using multiple access techniques and using multiple methods, 6. Distinguish the communication signal from all the background radiation from all the possible sources.

VLC leading researchers are working to solve these problems in the hardware of the transmitter-receiver itself[17], by designing compensating software[18] or using techniques, new materials and devices to reduce the consequences of such inconvenientes[19]. So far, within the perspective of new materials and devices, the efforts have being made mainly with traditional receivers in digital communications such as APD (Avalanche Photo Diode), PIN photo diode or photo diodes RGB[20]. However, to our knowledge, SiPM have not been considered for VLC. The SiPM are solid state photodetectors initially developed for applications in particle physics and recently in many other areas of R&D[21] as well for an increasing number of applications. This work deals with the study and the corresponding electronic design to profit of the outstanding characteristics of the SiPM to develop a VLC emitter-receiver metadevice for VLC. In this work we propose the SiPM as an alternative emitter-receiver for VLC supported on a full analysis of its functionality based on its special characteristics which are: very high sensitivity to be able to detect a single photon, very high gain of the order of $10^5$ and $10^6$, high speed response in the order of nanosecond, high bandwidth in the order of hundreds of MHz, immunity to magnetic fields and high PDE (Photon detection efficiency) [22]. The data obtained in the experiments and tests performed in this work, were organized, analyzed and presented to prove the direct consequences of the SiPM metadevice to solve the most of the drawbacks of VLC, increasing the practical applicability of this type of communication in more complex, uncertain and real scenarios than light-controlled and closed scenarios. The results also contribute to the solution of unsolved problems of the SiPM behavior such as frequency response, response time or transient state, direct response to fast and slow variations of background, saturation to high intensity of signal and background, and the processes *dark current*, *afterpulse* and *crosstalk*. Therese analysis of the VLC metadevice, performed with a well defined purpose experimental setup, allowed us to obtain a good characterization of the VLC metadevice as a promising emitter-receiver for VLC.

## 3. EXPERIMENTAL SETUP FOR A SIPM METASYSTEM AS VLC METADEVICE

The experimental platform for characterization of the frequency and time response of the SiPM as metasystem in radiation of high intensities, presented in Figure 1, consists of four key elements: 1. closed container, 2. Oscilloscope, 3. Signal Generator, and 4. voltage source. The closed container or black box, is for isolation of external light radiation, necessary for experimental observations with controlled light. The inner lining of the black box is aiming to generate as much noise as possible by isotropic white light reflected from everywhere inside the box; inside the black box two LEDs are used as radiation sources, one of them, LED2, for the emulation optical noise or background light and the other LED1, is located in front of the detector to generate pulses that emulate digital communication signal; both LEDs are reference UV5TZ-405-30 produced by the company VIBAR, and they have the same physical and electrical characteristics. Inside the black box there is a platform to accommodate the SiPM. The separation between the LED1 and the SiPM is 13.5 cm and both are aligned in order to capture as much as possible of the signal intensity in each test. The LEDs have a beam angle (solid angle of radiation) of 30 degrees where 50% of the power is emitted, the radiation wavelength is 405 nm ± 2.5 nm (purple), the operating voltage is 3.5 volts with a maximum current consumption of 15 mA. The LED2, is perpendicular to the line connecting the LED1 and the SiPM at a distance of 12 cm to produce background light from isotropic reflection in all the walls of the black box. As discussed in the description of the problem, there is a relationship between the transmission signal and the backlight, which is considered the background but no all the background noise, because there is electronic noise and thermal noise added in all the process from source to detection. This relationship must be identified from the observation data obtained by the experimental setup that will allow the analysis and identification of the effects of background on the SiPM response for a well defined and controlled transmission signal.

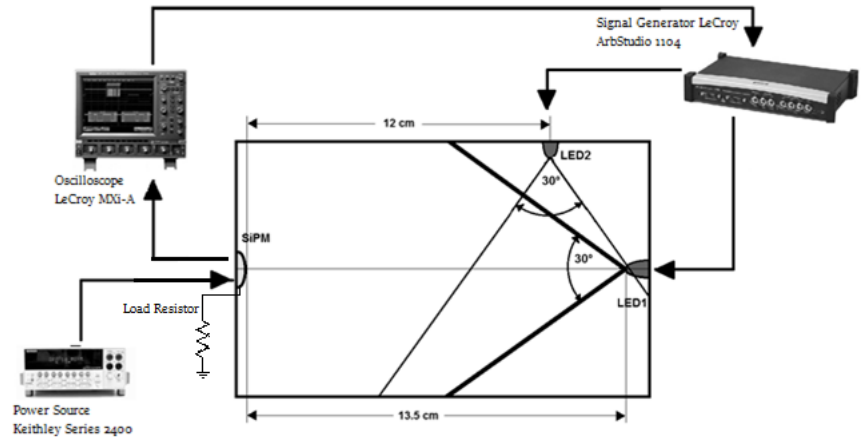

Figure 1. Experimental assembly. Representation of the four key elements of the experimental setup: The black box containing the SiPM and both LEDs, the oscilloscope to acquire the data, the generator to control the signals of LED1 and LED2, and power supplying operating the voltage source for the SiPM.

The oscilloscope is a LeCroy model MXi-A series, and the source is a Keithley series 2400. The oscilloscope capture the signal generated by the SiPM and makes measurements, processes and calculates various parameters of the signal in real time (harmonic decomposition, average signal strength, etc.), it also storages and converts the experimental data into digital format files.

The LeCroy generator lets the control the frequency and voltage applied to the radiation sources with a wide bandwidth (65 MHz) and a wide range of voltages (0-6 volts), values that emulate the real conditions of VLC, keeping a large bandwidth in data transmission and adequate voltage dynamic range to control the switching of radiation sources.

The experimental platform is designed to work at room temperature. Experiments were performed at 20.7 ± 0.6 ° C, with isolation of the outside light, ambient atmospheric pressure (560 mmHg) and polarization conditions consistent with the devices in use: SiPM S10365-11-025C with a bias voltage of -71.45 V and the LEDs UV5TZ-405-30 with operating voltage of 4V. The Keithley voltage source supplies a negative voltage to the polarization SiPM; this voltage generates reliable photo detector functionality.

As already mentioned, the SiPM photodetectors were initially developed for applications in particle physics and recently in various areas of knowledge and applications; they are capable of counting individual photons and are composed of an array of avalanche photodiodes in Geiger mode (reverse bias) or GM-APD, configured electrically in parallel. Each GM-APD operates in accordance with the internal photoelectric effect emitting an electric signal when it absorbs at least one photon. The electrical signal generated by the SiPM is the sum of the signals of all the GM-APD. The SiPM has a gain between $10^5$ to $10^6$ depending on the model[23].

The distribution of APDs in the SiPM conforms a grid where each APD represents a pixel; the SiPM used in this work has an area of 1 square millimeter and has 1600 (40x40 APDs) pixels, each pixel having an area of 625 square micrometers. In figure 2 we show a similar but smaller SiPM of 20x20 APDs. In figure 3 we observe the detection frequency (color of visible light) range of the SiPM, between 300 and 900 nanometers, with the maximum peak at 500 nm corresponding to violet[24].

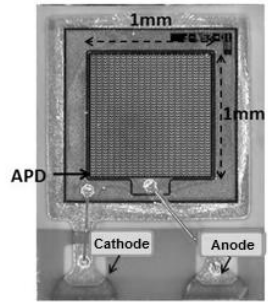

Figure 2. The matrix structure of the avalanche photodiode array of 20 x 20, in an area of 1 mm2 shown. The APD are connected in parallel, so that the SiPM having a single connector and one anode to cathode.

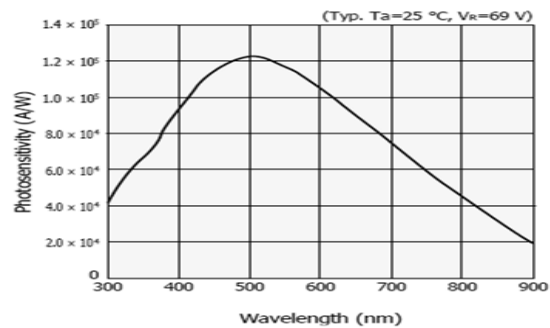

Figure 3. Spectral response S10365-11-025C SiPM Hamamatsu Photonics. The photosensitivity (A / W) vs. wavelength (λ) is shown. The vertical axis corresponds to the current produced by one watt of light detected. Source: Hamamatsu Photonics Datasheet[24].

It is well known that the performance of the SiPM and its gain strongly depend on temperature and bias voltage small variations[25]. However, there is still a strong need of better models in order to have a deeper understanding and control of these detectors and therefore improve the R&D and new applications of such high quality photo detectors. To develop such models, an important experimental effort have to be done to obtain sufficient high quality data of the characteristics and performance of the SiPMs as a function of temperature and bias voltage variations, among other important variables. Most of the related studies have been performed under conditions of weak radiation intensities corresponding to several or even a few hundreds of photons arriving to the APDs of the SiPM, while in VLC the fluxes of radiations correspond to millions of photons in each pulse of communication, forcing the SiPM to work in a continuous supersaturation[26]. The main problem for VLC is that the variation of the digital communication signal with respect to the background is relatively weak, therefore the great qualities of SiPM, in this case the high sensitivity in particular, provide the best alternative for the detection of these small intensity variations. The SiPM has the sufficient sensitivity to detect and differentiate small variations of intensity, corresponding to the signal, immersed in a high intensity of backlight, and transform them into an electrical signals of several millivolts. This small but detectable signal is amplified and transformed into a digital signal by the appropriated electronics that completes the VLC metadevice.

The LED used as the light source is composed of InGaN/sapphire and has a spectral response described in Figure 4. Each curve A, B, C and D corresponds to a different type of LED, the one used in this work corresponds to curve D that has its spectral peak at 405 nanometers. This LED was chosen because of their spectral similarity with the response of SiPM, in order to get the most out of the qualities of the SiPM.

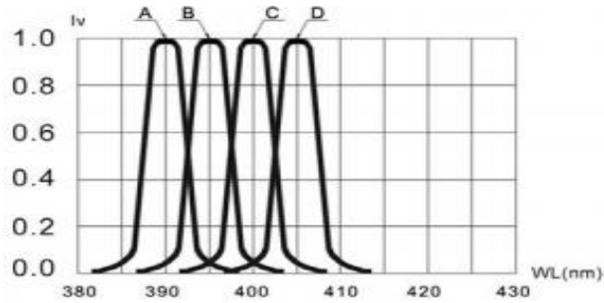

Figure 4. Relative Luminous Intensity vs. Wavelength. Spectral Response LED-405-30 UV5TZ represented by D. The curves A, B and C correspond to other LEDs not used in this work. Source: BIVAR datasheet[27].

Prior to this experimental setup, the characterization of the radiation sources, LEDS 1 and 2, by the radiation intensity as a function of the bias voltage, is presented in Figure 5. It was performed in order to establish a pattern of change in intensity measured with a light meter in units of LUX (units for illumination intensity of the international system). The light meter was located in the positioning of the SiPM as indicated in figure 1. In addition, a characterization of frequency response was performed for each radiation source, this is done in order to control the radiation intensity as a function of the frequency switching of the light sources. The light meter used for these measurements was an Amprobe LM-120 and the quantified relation obtained is presented in the plot of figure 5. This characterization of radiation sources allows a more precise control of experimental conditions and ensures the repeatability of the tests.

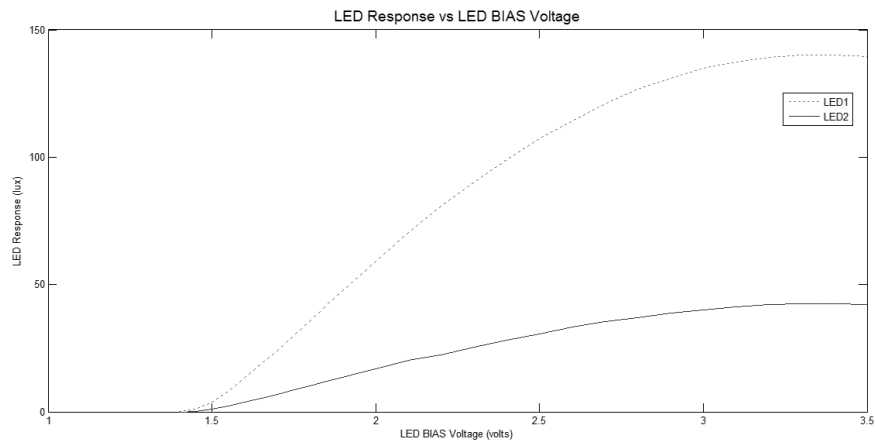

Figure 5. Results of LEDs characterization: response of LED1 and LED2 used as the light sources for communication signal and background respectively. Both LEDs are equal, exactly the same specifications.

## 4. RESULTS ANALYSIS

### 4.1 SiPM response to commutation frequency (without background)

The response of the SiPM to stimulus commutation frequency is presented in figure 6. SiPM. Voltage signal obtained from the SiPM is in the order of mV can be considered to present little variation (voltage is kept above 120mV) for commutation frequencies lower than 40 MHz. For higher frequencies, SiPM response decreases up to 50% of initial value (60mV) at 60MH as shown in figure 6. In communication applications, Signal to noise ratio is an important issue to determine the feasibility of the system. The electromagnetic noise is produced by many sources of natural and human origin and electronic devices must be robust enough to handle with it, as stated in the electromagnetic Compliance standard[28]. For our tests, the electromagnetic noise was identified to be in the order of 40mV, which compared to signal magnitude in the frequency range of 0 to 60 MHz, is small enough to be easily isolated from the communication signal. The results obtained, show that SiPM based VLC metadevice is feasible without further signal conditioning in the range of 0 to 47 MHz and that operational range can be extended with little signal conditioning effort.

The SiPM frequency response identified in this work, will be an important input to conceive the amplification and signal conditioning electronics for the VLC metadevice.

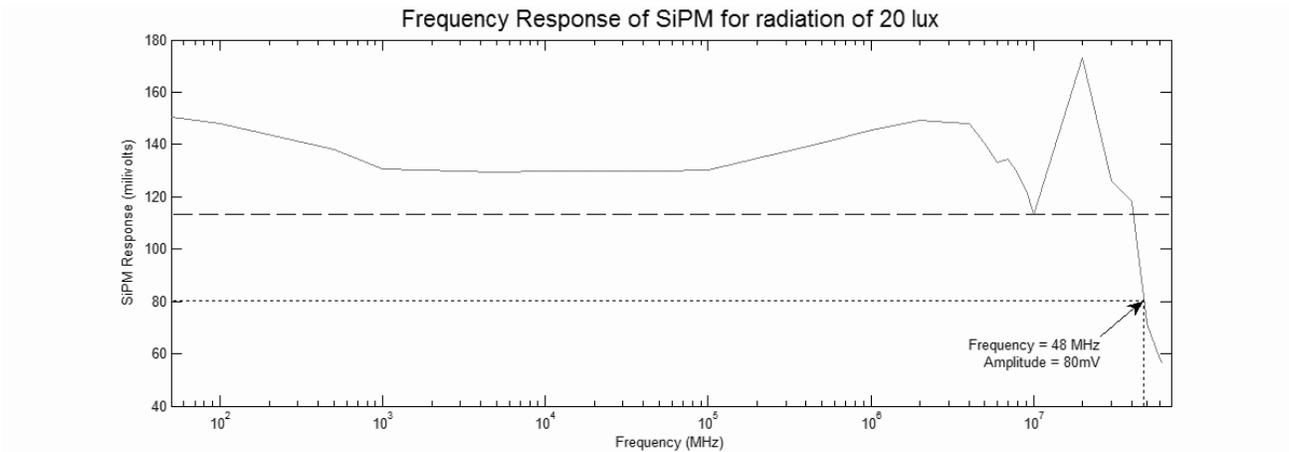

Figure 6. Commutation Frequency Response of SiPM for Radiation of 20 lx. SiPM Response (mV) vs Frequency (MHz).

### 4.2 SIPM TIME RESPONSE (WITHOUT BACKGROUND)

Tests for time response of SiPM were performed at: 21.3 °C, LED1 commutation frequency fixed in 1 KHz with a radiation intensity measures at SiPM of 140 lx, and a polarization voltage 3.4 volts, signal pulse width of 700 μs, LED2 is kept in "Off" state. The experimental parameters considered in tests to characterize the behavior of SiPM without background were: 1-rise time, 2- fall time, 3-stabilization time (Vs), as shown in figure 7.

Tests were performed six times with 10000 samples schema each. Results of tests are given in Table 1.

Table 1. Mean values for Rise time, Fall time and stabilization voltage (Vs).

| Parameter test | Rise time (ns) | Fall time (μs) | Vs (mV) |
|---|---:|---:|---:|
| **Repetition 1** | 2.475 | 2.396 | 167.430 |
| **Repetition 2** | 2.463 | 2.500 | 168.238 |
| **Repetition 3** | 2.050 | 2.406 | 166.530 |
| **Repetition 4** | 2.150 | 2.475 | 165.338 |
| **Repetition 5** | 2.506 | 2.438 | 165.919 |
| **Repetition 6** | 1.755 | 2.431 | 163.913 |
| **Mean** | 2.185±0.302 | 2.441±0.040 | 166.228±1.538 |

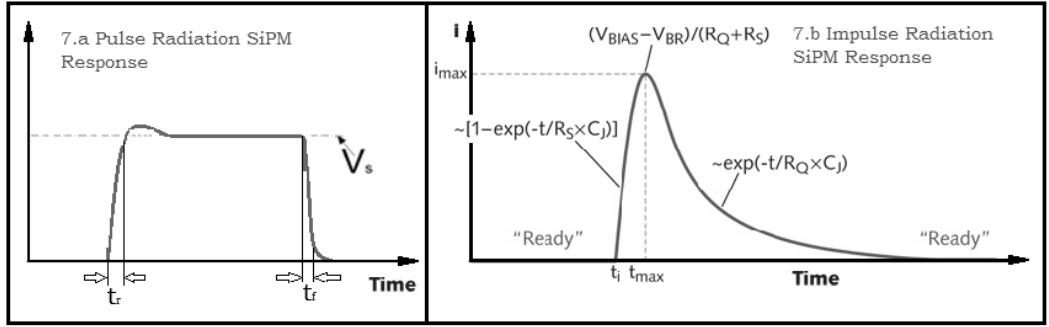

Figure 7. Parameters included in time response analysis for SiPM. Rise time (tr), fall time (tf) steady state voltage (Vs) are shown in 7. a. 7. b shows impulse response peak rise and decaying behavior.

### 4.3 SiPM-DC analysis (with background)

For a DC response analysis of the SiPM, the experimental conditions were established and recorded maintaining a temperature of 20.1 ° C, resistance load of 49.49 Ω SiPM, and a step signal of luminous intensity supplied by the LED2 as indicated by figure 8. The initial continuous radiation intensity was set at 2.2 lx with increments of 2 lx up to 42.2. For each step we took 1000 samples of the SiPM response.

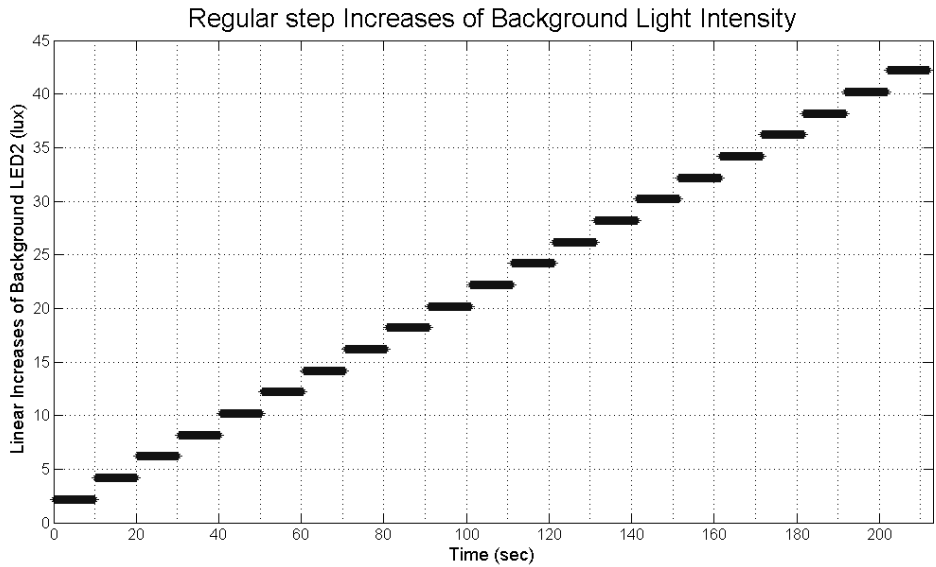

Figure 8. Background radiation signal from LED2 used for the analysis of the SiPM response to increasing intensity.

The response of the SiPM to constant radiation is used to illustrate the relation of the luminous stimulus with the polarization voltage as illustrated in Figure 9. Since the SiPM is connected in series with the load resistor as shown in Figure 1, and the voltage supplied by the power source is constant

$$V_s = V_{BIAS} + V_{LOAD} \qquad (1)$$

the voltage on the load resistance varies as shown in Figure 9. The load resistor limits the inrush current to strike the SiPM bias voltage according to the equation

$$Q = C(V_{BIAS} - V_{BD}) \qquad (2)$$

where C is the equivalent capacitance of the SiPM, normally of the order of pF.

Since

$$V_{LOAD} = I_{av} R_{LOAD} \qquad (3)$$

from (1), (2) y (3)

$$Q = C(V_s - I_{av} R_{LOAD} - V_{BD}) \qquad (4)$$

we obtain

$$I_{av} = \frac{-(\frac{Q}{C} - V_s + V_{BD})}{R_{LOAD}} \qquad (5)$$

This behavior is important in photodetection phenomena in low light intensities where only a few photons striking the SiPM can be taken as discrete events. Considering that the SiPM time constant is smaller than the time between two events, allows the relaxation of the SiPM and the recovery of the initial conditions, resulting in an impulse response. A continuous radiation, where for the given time scale striking photon events can no longer be considered as discrete, prevents SiPM to achieve a complete relaxation between events; this phenomena produces a shift of the bias voltage, that can be experimentally observed in figure 9 as a decreasing bias voltage with increasing radiation intensity; In this scenario, as light intensity increases, the bias voltage variation is constraint to a smaller range, resulting in an attenuated response of the SiPM. An initial approach to this phenomenon is made by Grubber et. al[29].

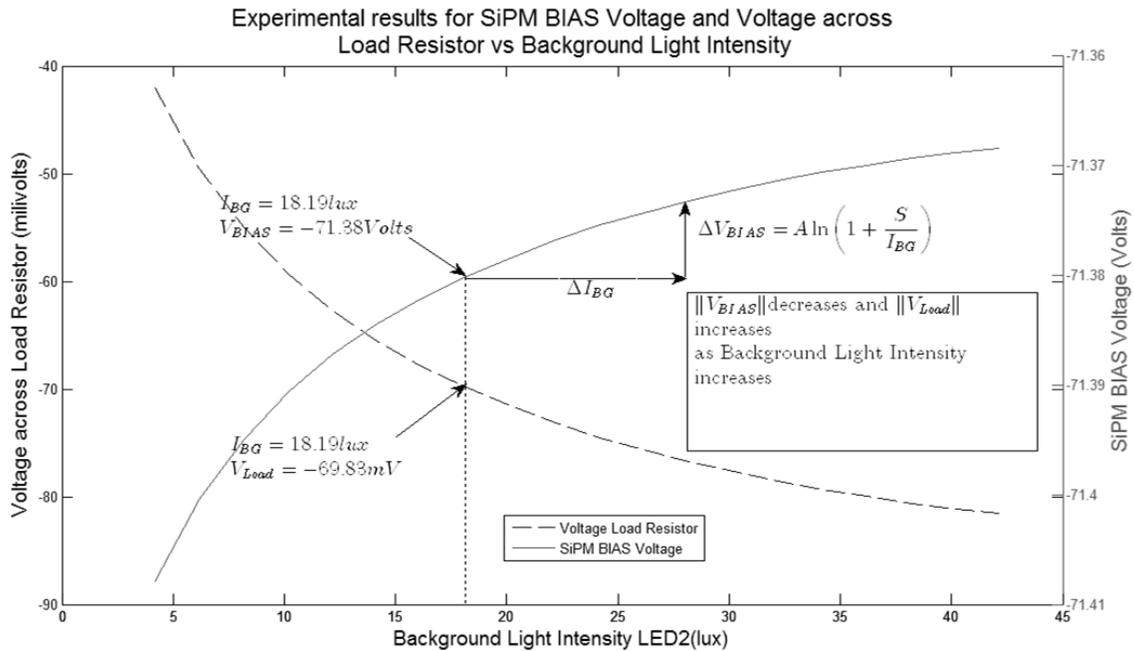

Figure 9. Experimental results for SiPM Bias Voltage and Voltage across Resistor vs Background Light Intensity.

When the SiPM is polarized near the breakdown voltage (reverse voltage necessary for the arrangement of APDs to start charge conduction in reverse), it generates a very weak inverse current that increase in proportion to the radiation increase. In fact, the voltage variation in the load resistance results from the variation of the inverse current driving behavior explained by Ohm's Law. Variations of the inverse current show a non-linear behavior of logarithmic nature.

The reverse bias current varies with the intensity of light radiation with a logarithmic relationship; hence, from equations (1) and (3) it can be deduced that the variation of the bias voltage ($V_{BIAS}$), with respect to the intensity of light radiation, is also logarithmic, as shown in Figure 9.

Defining $V_{out}$ as the amplitude of the response signal of the SiPM to a light intensity signal S, in the presence of background radiation intensity $I_{BG}$, we have:

$$V_{out} = A\left[\ln(1 + \frac{S}{I_{BG}})\right] \qquad (6)$$

Defining the gain of the metadevice as

$$G = \frac{V_{out}}{S} \qquad (7)$$

We obtain the gain as a two-parameter function of the form

$$G(I_{BG}, S) = \frac{A[\ln(I_{BG}+S)]}{S} - \frac{A\ln(I_{BG})}{S} \qquad (8a)$$

$$G(I_{BG}, S) = \frac{A\left[\ln(1+\frac{S}{I_{BG}})\right]}{S} \qquad (8b)$$

For an operating range OIR (Operational Intensity Range) of the signal S, defined by

$$0 < S \leq Imaxsat - I_{BG} \qquad (9)$$

where *Imaxsat* defines a maximum value of gain technologically feasible corresponding to the SiPM in use.

Assuming that a signal conditioning is possible for a signal to noise ratio SNR:

$$SNR > k \text{ for } SNR = \frac{S}{N} \qquad (10)$$

where N is the size of the average amplitude of the noise signal and k is a criterion of technological capabilities of signal discrimination, we can verify that:

$$V_{out} > kN \qquad (11)$$

From equations (6) and (10), we obtain

$$V_{out} = A\left[\ln(1 + \frac{S}{I_{BG}})\right] > kN \qquad (12)$$

and from (12) we obtain the relation

$$\frac{S}{I_{BG}} > e^{\frac{kN}{A}} - 1 \qquad (13)$$

showing that for a given radiation level of background $I_{BG}$, the magnitude of the communication signal is limited by the background signal strength and technological capabilities of discrimination, as the following equation indicates:

$$S > (e^{\frac{kN}{A}} - 1)I_{BG} \qquad (14)$$

Seeing it from another point of view, given a communication signal strength S, the metadevice limit of effective operation with a background intensity $I_{BG}$, is defined by

$$I_{BG} < \frac{S}{(e^{\frac{kN}{A}} - 1)} \qquad (15),$$

given the restriction of equation (9).

As can be deduced from the nonlinear relationship shown in (8a) and (8b), an analog signal will be affected in its form factor, which can affect the signal necessary conditions for data transmission, according to the type of modulation used.

As already mentioned, the breakdown voltage of the SiPM is one of the determining factors in its response and has a high dependence on temperature[30], affecting the bias voltage. The study in this article only considers the gain variations due to changes in light intensity, the influence of temperature changes on the gain of the metadevice is subject of research in progress.

The study of the SiPM as a central component of the VLC metadevice, have been performed under a maximum light intensity of 140 lx, to which the VLC metadevice had a response of 160 millivolt signals before amplification. This signal large enough to be used in a communication system. Even dough experimental setup did not allow to perform tests with higher light intensity, in order to identify Imaxsat and thus the metadevice operational limit, experimental results provide evidence that the VLC metadevice is suitable for light communications, since 140 lx corresponds to a soft room artificial lighting, a real environmental condition for VLC.

### 4.4 Signal Amplitude vs Background

As expected, background has negative effects on the transmission signal, as shown in equation (8). The background attenuates and may distort the SiPM response to the incident information signal. Figure 10 shows the attenuation of the SiPM response increasing the background light intensity by steps of 2 lx. LED2 is used as radiation source for this test to emulate the background, while LED1 emulates the transmission signal. The frequency of the transmission signal (LED1) is fixed at 100 Hz, the voltage at 3.4 volts, with a pulse of 5 ms and the temperature at 20.4 ° C. The initial radiation value of the LED2 is set at 2.2lx and increased by steps of 2 lx up to 42.2 lx.

The attenuation of the SiPM response because of background shows a logarithmic behavior decreasing as the background amplitude increases, as shown in Figure 10. We observe a bilateral attenuation made of a background component and an amplitude component, as indicated in equation (8a). Experimentally, a background amplitude of 4.2 lx reduces the SiPM response by 50% of the response without background,

Given the logarithmic nature of equation (8b) we can obtain the expression

$$\lim_{I_{BG} \to \infty} \frac{dG}{dI_{BG}} = 0 \qquad (16)$$

implying that the greatest variations of attenuation are obtained at low intensities, being less significant at high intensities. This feature is important, because the impact of the background on the gain is limited by the logarithmic function representing the robustness of the VLC metadevice at high intensities of background. As mentioned before, the background noise and signal may be discriminated with an appropriated design of an electronic system that consider, and take advantage of the special characteristics of the SiPM and the communication requirements to constitute the VLC metadevice. Precisely, by knowing and measuring with precision the mentioned effects, it is possible to design the mentioned electronic system for VLC, therefore, the electronics and the SiPM altogether as a metadevice, offers significant advantages in the transmission of information through visible light.

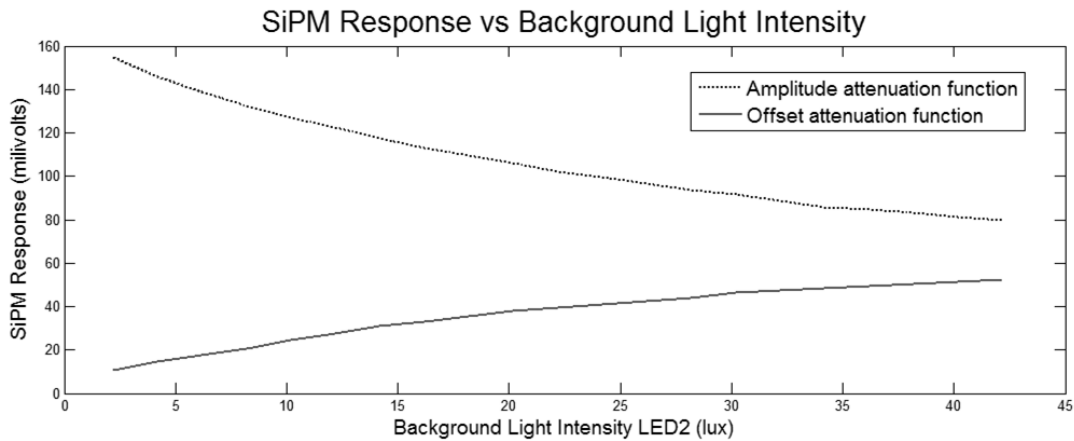

Figure 10. Attenuation of the SiPM bilateral response for increasing background intensity. When background intensity increases the amplitude of the response decreases logarithmically. The effect is visible both in the upper and the lower amplitude, shaping these envelope curves.

As it was presented, it is feasible to meet the requirements of VLC in real conditions of background with the present state of art electronics and photomultipliers.

## 5. METADEVICE DESCRIPTION

The VLC metadevice conceived in this work follows the structure of other communication devices, integrating the SiPM as receiver. At the actual state, the metadevice is capable of transferring digital and processing digital signals as shown in figure 11. The study of the behavior of the SiPM as a metasystem has generated the knowledge and precise specifications to design a hybrid electronic system (analog and digital) capable of processing the signal of a SiPM to meet the requirements for an optimal VLC. The conditions are: Bandwidth (BW), Slew Rate, Rising and Falling times, voltage stabilization, signal background discriminator, gain, and CMRR (Common Mode Rejection Ratio). These specifications are derived from data obtained in the experimental tests described above. The VLC metadevice Bandwidth is determined by the transfer rate of the communication system, which for this work is fixed to 80 MHz for the developed application, implementing communication with On-Off Keying modulation. Under this conditions, a transfer rate of 160 Mbps was achieved. According to the measurements, the Rise Time and Fall Time are 2.2 ns and 2.4 ms respectively and a voltage stabilization output equal to 3.3 volts for an input voltage of 160 millivolts in each pulse. The calculated Slew Rate to a pulse with amplitude of 160 mV, was found to be of 0.8 V/ms. The slew rate of electronics performing signal conditioning is higher than the one observed for the SiPM, this fact allows the metadevice to adequate processing SiPM signals while giving room to metadevice improvement as new SiPM may have quicker responses. The developed discriminator of background isolates the effect of the backlight on the response of the SiPM. The gain should be sufficient to generate a pulse amplitude of 3.3 volts from a 160 mV signal, in this case the gain is different and is part of the design of the specific electronic system for this application. The CMRR is included to reject any electromagnetic noise present simultaneously in all the system inputs.

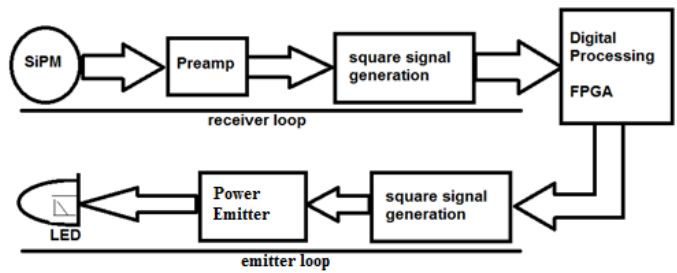

Figure 11. Block diagram of the VLC transmitter-receiver metadevice. The "square signal generation" block, in the receiver loop, indicates the generation of a square wave from the information signal received from the preamplifier which in turn amplifies the response SiPM to light signal. The same block in the emitter loop generates the square wave transmitter to feed the power stage lighting for the transmission. The "Digital Processing FPGA" block, is the specific purpose electronic, designed to process and deliver the information received to a computer, and also to generate the information to be transmitted by the LED.

## 6. CONCLUSIONS

It is experimentally demonstrated that the SiPM used in saturation mode, can be an excellent receiver for a VLC metadevice, improving several characteristics of traditional VLC receivers and solving some of the limitations of light communication systems. This conclusion is obtained from the SiPM characterization in VLC conditions: response times, gain, and robustness to background and sensitivity having comparative advantages over the photo detectors used today.

The natural background present in a real environment VLC, has an important impact on the gain of the SiPM but, because of the special qualities of such photomultiplier, it is not an obstacle for the proposed VLC metadevice, on the contrary, it is a problem exceptionally well solved by the use of the SiPM compared with other photo detectors traditionally used in VLC. This conclusion follows from equation (6) and the results obtained with a background of 140 lx; in these conditions the signal presents a gain large enough to be detected and discriminated from coexisting noise fulfilling the minimum requirements of a VLC system. We also found that the Vout is about 100 mV, much higher than the standard voltage magnitudes easily handled by commercial electronics and, due to the high sensitivity of the SiPM, the VLC metadevice can reconstruct signals with frequencies of the order of 70 MHz, sufficient for VLC systems using different modulation types.

## 7. ACKNOWLEDGMENT

We would like to thank the support of the CICBA and DCA of the UAN for their resources and facilities dedicated to this work and the continuous encouragement. We thank the collaboration and useful discussions with doctors Harold Yepes, Carlos Montaño and the technical support of Jhon Edward Cruz.